\documentclass[twoside,fleqn,espcrc2]{elsarticle}
\usepackage{lineno,hyperref}
\modulolinenumbers[5]

\usepackage[utf8]{inputenc}
\usepackage{subcaption}
\usepackage{multirow}
\usepackage{bm}
\usepackage{verbatim}
\usepackage{color}
\usepackage[table,xcdraw]{xcolor}
\usepackage{booktabs}
\usepackage[numbers]{natbib}
\usepackage{graphicx}
\usepackage{amssymb}
\usepackage{amsmath}
\usepackage{mathabx}
\usepackage{soul}
\usepackage{hhline}
\usepackage{epstopdf}

\usepackage{enumitem}
\usepackage[font=small,skip=2pt]{caption}
\graphicspath{{Figures/}}

\newcolumntype{L}[1]{>{\raggedright\let\newline\\\arraybackslash\hspace{0pt}}m{#1}}
\newcolumntype{C}[1]{>{\centering\let\newline\\\arraybackslash\hspace{0pt}}m{#1}}
\newcolumntype{R}[1]{>{\raggedleft\let\newline\\\arraybackslash\hspace{0pt}}m{#1}}

\usepackage{ifthen}
\makeatletter
\renewcommand\@cite[2]{%
Ref.~#1\ifthenelse{\boolean{@tempswa}}
{, \nolinebreak[3] #2}{}
}
\renewcommand\@biblabel[1]{#1.}

\makeatother
\usepackage[textwidth=3.7cm]{todonotes}

\makeatletter
\def\ps@pprintTitle{%
\let\@oddhead\@empty
\let\@evenhead\@empty
\def\@oddfoot{}%
\let\@evenfoot\@oddfoot}
\makeatother

\usepackage{todonotes}

\newcommand{\modified}[1]{#1}
\newcommand{\revision}[1]{#1}

\begin{document}
    \begin{frontmatter}
        \title{Boundary loss for highly unbalanced segmentation}

        \author[LIVIA]{Hoel Kervadec\corref{mycorrespondingauthor}}
        \author[LIVIA]{Jihene Bouchtiba}
        \author[LIVIA]{Christian Desrosiers}
        \author[LIVIA]{Eric Granger}
        \author[LIVIA]{Jose Dolz}
        \author[LIVIA,CRCHUM]{Ismail Ben Ayed}

        \address[LIVIA]{\'ETS Montréal, Canada}
        \address[CRCHUM]{CRCHUM (University of Montreal Hospital Centre), Canada}

        \cortext[mycorrespondingauthor]{Corresponding author: hoel@kervadec.science}

        \begin{abstract}
            Widely used loss functions for CNN segmentation, e.g., Dice or cross-entropy, are based on integrals over the segmentation regions. Unfortunately, for
            highly unbalanced segmentations, such regional summations have values that differ by several orders of magnitude across classes, which affects training performance and stability. We propose a {\em boundary} loss, which takes the form of a distance metric on the space of contours, not regions. This can mitigate the difficulties
            of highly unbalanced problems because it uses integrals over the interface between regions instead of unbalanced integrals over the regions. Furthermore, a boundary loss complements regional information.
            Inspired by graph-based optimization techniques for computing active-contour flows,
            we express a non-symmetric $L_2$ distance on the space of contours as a regional integral, which avoids completely local differential computations involving contour points. This yields a boundary loss expressed with the regional softmax probability outputs of the network, which can be easily combined with standard regional losses and implemented with any existing deep network architecture
            for N-D segmentation.  We report comprehensive evaluations and comparisons on different unbalanced problems, showing that our boundary loss can yield significant increases in performances while improving training stability. Our code is publicly available\footnote{\url{https://github.com/LIVIAETS/surface-loss}}.
        \end{abstract}

        % TL;DR: We propose a boundary loss based on L2 distance and evaluate it on two highly unbalanced segmentation problems.

    \end{frontmatter}
    % \linenumbers

    % \tableofcontents  % Only to debug and see the struct of the doc

    \section{Introduction}
\begin{figure}[t]
            \centering
            \begin{subfigure}[b]{0.3\textwidth}
                \includegraphics[width=\textwidth]{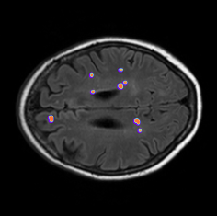}
                \caption{Ground truth}
            \end{subfigure}
            \begin{subfigure}[b]{0.3\textwidth}
                \includegraphics[width=\textwidth]{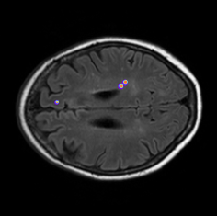}
                \caption{GDL}
            \end{subfigure}
            \begin{subfigure}[b]{0.3\textwidth}
                \includegraphics[width=\textwidth]{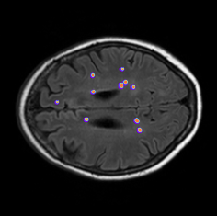}
                \caption{GDL + boundary loss}
            \end{subfigure}
            \caption{A visual comparison that shows the positive effect of our boundary loss on a validation data from the WMH dataset. Our boundary loss helped to recover small regions that were otherwise missed by \revision{the model trained with} the generalized Dice loss (GDL). Best viewed in colors.}
            \label{fig:gt_gdl_surface}
        \end{figure}

        Recent years have witnessed a substantial growth in the number of deep learning methods for medical image segmentation \cite{litjens2017survey,shen2017deep,dolz20173d,ker2018deep}. Widely used loss functions for segmentation, e.g., Dice or cross-entropy, are based on {\em regional} integrals, which are convenient for training deep neural networks. In practice, these regional integrals are summations over the segmentation regions of differentiable functions, each directly invoking the softmax probability outputs of the network. Therefore, standard stochastic optimizers such \revision{as} SGD are directly applicable. Unfortunately, difficulties occur for highly unbalanced segmentations, for instance, when the size of target foreground region is several orders of magnitude less than the background size. For example, in the characterization of white matter hyperintensities (WMH) of presumed vascular origin, the foreground composed of WMH regions may be 500 times smaller than the background (see the typical example in Fig. \ref{fig:gt_gdl_surface}). In such cases, quite common in medical image analysis, standard regional losses contain foreground and background terms with values that differ considerably, typically by several orders of magnitude, potentially affecting performance and training stability \cite{milletari2016v,sudre2017generalised}.

        Segmentation approaches based on convolutional neural networks (CNN) are typically trained by minimizing the cross-entropy (CE), which measures an affinity between the regions defined by probability softmax outputs of the network and the corresponding ground-truth regions. The standard regional CE has well-known drawbacks in the context of highly unbalanced problems. It assumes identical importance distribution of all the samples and classes. To achieve good generalization, it requires a large training set with balanced classes. For unbalanced data, CE typically results in unstable training and leads to decision boundaries biased towards the majority classes. Class-imbalanced learning aims to mitigate learning bias by promoting the importance of infrequent labels. In medical image segmentation, a common strategy is to re-balance class prior distributions by down-sampling frequent labels \cite{havaei2017brain,valverde2017improving}. Nevertheless, this strategy limits the information of the images used for training. Another common practice is to assign weights to the different classes that are inversely proportional to the frequency of the corresponding labels \cite{brosch2015deep,ronneberger2015u,kamnitsas2017efficient,long2015fully,yu2017volumetric}. In this scenario, the standard cross-entropy (CE) loss is modified so as to assign more importance to the rare labels. Although effective for some unbalanced problems, such weighting methods may undergo serious difficulties when dealing with highly unbalanced datasets, as seen with WMH segmentation. The CE gradient computed over the few pixels of infrequent labels is typically noisy, and amplifying this noise with a high class weight may lead to instability.

        The well-known Dice overlap coefficient was also adopted as a regional loss function, typically outperforming CE in unbalanced medical image segmentation problems \cite{milletari2016v,milletari2017hough,wong20183d}. Sudre et al. \cite{sudre2017generalised} generalized the Dice loss \cite{milletari2016v} by weighting according to the \revision{squared} inverse of class-label frequency. Despite these improvements over CE \cite{milletari2016v,sudre2017generalised}, regional Dice losses may encounter difficulties when dealing with very small structures. In such highly unbalanced scenarios, mis-classified pixels may lead to large decreases of the loss, resulting in unstable optimization. Furthermore, Dice corresponds to the harmonic mean between precision and recall, implicitly using the arithmetic mean of false positives and false negatives. False positives and false negatives are, therefore, equally important when the true positives remain the same, making this loss mainly appropriate when both types of errors are equally high. The recent research in \cite{salehi2017tversky,abraham2018novel} investigated losses based on the Tversky similarity index in order to provide a better trade-off between precision and recall. It introduced two parameters that control the importance of false positives and false negatives. Other recent advances in class-imbalanced learning for computer vision problems have been adopted in medical image segmentation. For example, inspired by the concept of focal loss \cite{lin2018focal}, Dice and Tvserky losses have been extended to integrate a focal term, which is parameterized by a value that controls the importance of easy and hard training samples \cite{abraham2018novel,wong20183d}. \revision{Furthermore, the combination of several of these regional losses has been further investigated \cite{zhu2019anatomynet}.} The main objective of these losses is to balance the classes not only in terms of their relative class sizes, but also by the level of segmentation difficulty.

        \revision{More recently, Karimi et al. \cite{karimi2019reducing} proposed a novel loss function that attempts to directly reduce the Hausdorff distance (HD). This relaxed loss based on the HD is shown to bring improvements when combined with the DSC loss. Nevertheless, its main drawback is the high computational cost of computing the distance transforms. Particularly, at each training epoch, the new distance maps have to be recomputed for all the images, which incurs in a computationally costly process. This issue is further magnified in the case of 3D volumes, which heavily increases the computational burden.}

        \subsection*{Contributions}
            In this paper, we propose a {\em boundary} loss that takes the form of a distance metric on the space of contours (or shapes), not regions. We argue that a boundary loss can mitigate the issues related to regional losses in highly unbalanced segmentation problems. Rather than using unbalanced integrals over the regions, a boundary loss uses integrals over the boundary (interface) between the regions. Furthermore, it provides information that is complementary to regional losses. It is, however, challenging to represent the boundary points corresponding to the regional softmax outputs of a CNN. This difficulty may explain why boundary losses have been avoided in the context of deep segmentation networks. Our boundary loss is inspired by techniques in discrete graph-based optimization for computing gradient flows of curve evolution \cite{boykov2006integral}. Following an integral approach for computing boundary variations, we express a non-symmetric $L_2$ distance on the space of shapes (or contours) as a regional integral, which avoids completely local differential computations involving contour points. This yields a boundary loss expressed as the sum of linear functions of the regional softmax probability outputs of the network. Therefore, it can be easily combined with standard regional losses and implemented with any existing deep network architecture for N-D segmentation.

            We evaluated our boundary loss in conjunction with various region-based losses
            on two challenging and highly unbalanced segmentation problems -- the Ischemic Stroke Lesion (ISLES) and the White Matter Hyperintensities (WMH) benchmark datasets. The results indicate that the proposed boundary loss yields a more stable learning process, and can bring significant gains in performances, in terms of Dice and Hausdorff scores.

    \section{Formulation}
        Let $I: \Omega \subset {\mathbb R}^{2,3}  \rightarrow {\mathbb R}$ denotes a training image with spatial domain $\Omega$, and $g:\Omega \rightarrow \{0,1\}$ a binary ground-truth segmentation of the image: $g(p) = 1$ if pixel/voxel $p$ belongs to the target region $G \subset \Omega$ (foreground region) and $0$ otherwise, i.e., $p \in \Omega \setminus G$ (background region)\footnote{We focus on two-region segmentation to simplify the presentation. However, our formulation extends to the multi-region case in a straightforward manner.}. Let $s_\theta: \Omega \rightarrow [0,1]$ denotes the softmax probability output of a deep segmentation network, and $S_{\theta} \subset \Omega$ the corresponding segmentation region: $S_{\theta} = \{p \in \Omega \, | \, s_{\theta}(p) \geq \delta \}$ for some threshold $\delta$. Widely used segmentation loss functions involve a {\em regional integral} for each segmentation region in $\Omega$, which measures some similarity (or overlap) between the region defined by the probability outputs of the network and the corresponding ground-truth. In the two-region case, we have an integral of the general form $\int_{\Omega} g(p) f(s_{\theta}(p))dp$ for the foreground, and  of the form $\int_{\Omega} (1-g(p))f(1-s_{\theta}(p))dp$ for the background. For instance, the standard two-region cross-entropy loss corresponds to a summation of these two terms for $f=-\log(\cdot)$. Similarly, the generalized Dice loss (GDL) \cite{sudre2017generalised} involves regional integrals with $f=1$, subject to some normalization, and is given as follows for the two-region case:
        \begin{equation}
            \label{eq:gdl}
            {\cal L}_{GDL}(\theta) = 1 - 2\frac{w_G \int_{p \in \Omega} g(p) s_{\theta}(p)dp + w_B \int_{p \in \Omega} (1-g(p))(1-s_{\theta}(p))dp}{w_G \int_{\Omega} [s_\theta(p) + g(p)]dp + w_B \int_{\Omega} [2-s_\theta(p) - g(p)]dp}
        \end{equation}
        where coefficients $w_G = 1/\left ( \int_{p\in \Omega} g(p)dp \right)^2$ and $w_B = 1/\left ( \int_{\Omega} (1 - g(p))dp \right )^2$ are introduced to reduce the well-known correlation between the Dice overlap and region size.

        Regional integrals are widely used because they are convenient for training deep segmentation networks. In practice, these regional integrals are summations of differentiable functions, each invoking directly the softmax probability outputs of the network, $s_{\theta}(p)$. Therefore, standard stochastic optimizers such SGD are directly applicable. Unfortunately, extremely unbalanced segmentations are quite common in medical image analysis, where, e.g., the size of the target foreground region is several orders of magnitude smaller than the background size. This represents challenging cases because the foreground and background terms have substantial differences in their values, which affects segmentation performance and training stability \cite{milletari2016v,sudre2017generalised}.

        Our purpose is to build a boundary loss $\mbox{Dist}(\partial G, \partial S_{\theta})$, which takes the form of a distance metric on the space of contours (or region boundaries) in $\Omega$, with $\partial G$ denoting a representation of the boundary of ground-truth region $G$ (e.g., the set of points of $G$, which have a spatial neighbor in background $\Omega \setminus G$) and $\partial S_{\theta}$ denoting the boundary of the segmentation region defined by the network output. On the one hand, a boundary loss should be able to mitigate the above-mentioned difficulties for unbalanced segmentations: rather than using unbalanced integrals within the regions, it uses integrals over the boundary (interface) between the regions. Furthermore, a boundary loss provides information that is different from and, therefore, complimentary to regional losses. On the other hand, it is not clear how to represent boundary points on $\partial S_{\theta}$ as a differentiable function of regional network outputs $s_\theta$. This difficulty might explain why boundary losses have been mostly
        avoided in the context of deep segmentation networks.

        \begin{figure}
            \centering
            \begin{subfigure}[b]{0.35\textwidth}
                \includegraphics[width=\textwidth]{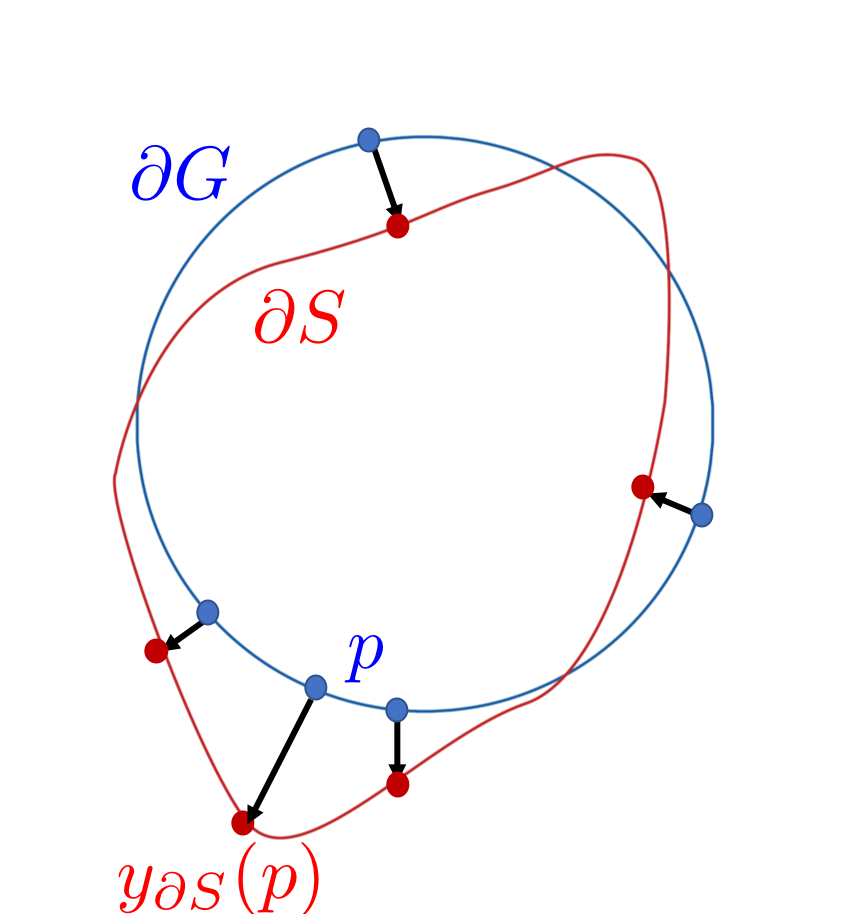}
                \caption{Differential}
            \end{subfigure}
            \begin{subfigure}[b]{0.35\textwidth}
                \includegraphics[width=\textwidth]{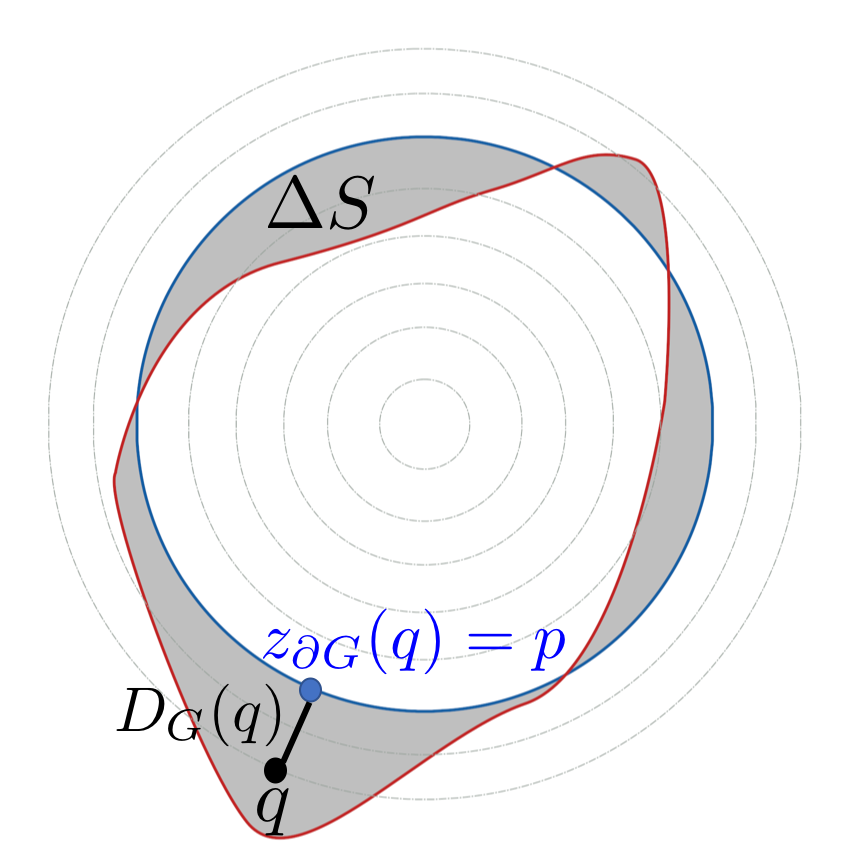}
                \caption{Integral}
            \end{subfigure}
            \caption{The relationship between {\em differential} and {\em integral} approaches for evaluating boundary change (variation).}
            \label{differential-vs-integral-approach}
        \end{figure}

        Our boundary loss is inspired from discrete (graph-based) optimization techniques for computing gradient flows of curve evolution \cite{boykov2006integral}. Similarly to our problem, curve evolution methods require a measure for evaluating boundary
        changes (or variations). Consider the following non-symmetric $L_2$ distance on the space of shapes, which evaluates
        the change between two nearby boundaries $\partial S$ and $\partial G$ \cite{boykov2006integral}:
        \begin{equation}
            \label{Differential-framework}
            \mbox{Dist}(\partial G, \partial S) = \int_{\partial  G}\| y_{\partial S}(p) - p\|^2 dp
        \end{equation}
        where $p \in \Omega$ is a point on boundary $\partial G$ and $y_{\partial S}(p)$ denotes the corresponding point on boundary $\partial S$, along the direction normal to $\partial G$, i.e., $y_{\partial S}(p)$ is the intersection of $\partial S$ and the line that is normal to $\partial G$ at $p$ (See Fig. \ref{differential-vs-integral-approach}.a for an illustration)
        $\|.\|$ denotes the $L_2$ norm. In fact, this {\em differential} framework for evaluating boundary change is in line with standard variational curve evolution methods \cite{mitiche-benayed-2011}, which compute the motion of each point $p$ on the evolving curve as a velocity along the normal to the curve at point $p$. Similarly to any contour distance invoking directly points on contour $\partial S$, expression \eqref{Differential-framework} cannot be used directly as a loss for $\partial S = \partial S_{\theta}$. However, it is easy to show that the differential boundary variation in \eqref{Differential-framework} can be approximated using an {\em integral} approach \cite{boykov2006integral}, which avoids completely local differential computations involving contour points and represents boundary change as a regional integral:
        \begin{equation}
            \label{integral-framework}
            \mbox{Dist}(\partial G, \partial S) \approx  2 \int_{\Delta S}D_G(q)dq
        \end{equation}
        where ${\Delta S}$ denotes the region between the two contours
        % \footnote{$\Delta S = (S \cup G) \setminus (S \cap G)$}
        and $D_G: \Omega  \rightarrow {\mathbb R}^+$ is a {\em distance map} with respect to boundary $\partial G$, i.e.,
        $D_G(q)$ evaluates the distance between point
        $q \in \Omega$ and the nearest point $z_{\partial G}(q)$ on contour $\partial G$: $D_G(q) = \|q - z_{\partial G}(q)\|$.
        Fig. \ref{differential-vs-integral-approach}.b illustrates
        this integral framework for evaluating the boundary distance in Eq. \eqref{Differential-framework}. To clarify approximation \eqref{integral-framework}, notice that integrating the distance map $2D_G(q)$ over
        the normal segment connecting a point $p$ on $\partial G$ and $y_{\partial S}(p)$ yields $\|y_{\partial S}(p) - p\|^2$, via
        the following variable change:
        \[\int_{p}^{y_{\partial S}(p)} 2 D_G(q)dq = \int_{0}^{\|y_{\partial S}(p) - p\|} 2 D_G d D_G =  \|y_{\partial S}(p) - p\|^2\]
        Thus, from approximation \eqref{integral-framework}, the non-symmetric $L_2$ distance between contours in Eq. \eqref{Differential-framework} can be expressed as a sum of regional integrals based on a {\em level set} representation of boundary $\partial G$:
        \begin{equation}
            \label{integral-binary}
        \frac{1}{2} \mbox{Dist}(\partial G, \partial S)  =  \int_{S} \phi_G(q)dq - \int_{G} \phi_G(q)dq = \int_{\Omega} \phi_G(q) s(q) dq - \int_{\Omega} \phi_G(q) g(q)dq
        \end{equation}
        where $s:\Omega \rightarrow \{0,1\}$ is binary indicator function of region $S$: $s(q) = 1$ if $q \in S$ belongs to the target and $0$ otherwise.
        $\phi_G: \Omega  \rightarrow {\mathbb R}$
        denotes the level set representation of boundary $\partial G$:
        $\phi_G(q) = - D_G(q)$ if $q \in G$ and $\phi_G(q) = D_G(q)$ otherwise.
        Now, for $S=S_{\theta}$, i.e., replacing binary variables $s(q)$ in Eq. \eqref{integral-binary} by the softmax probability outputs of the network $s_{\theta}(q)$, we obtain the following boundary loss which, up to a constant independent of $\theta$, approximates boundary distance $\mbox{Dist}(\partial G, \partial S_{\theta})$:
        \begin{equation}
            \label{boundary-loss}
            {\cal L}_B (\theta) = \int_{\Omega} \phi_G(q) s_{\theta}(q) dq
        \end{equation}
        Notice that we omitted the last term in Eq. \eqref{integral-binary} as it is independent of network parameters.
        The level set function $\phi_G$ is pre-computed directly from the ground-truth region $G$. In practice, our boundary loss in Eq. \eqref{boundary-loss} is the sum of linear functions of the regional softmax probability outputs of the network. Therefore, it can be easily combined with standard regional losses ($\mathcal{L}_R$) and implemented with any existing deep network architecture for N-D segmentation\modified{:
            \begin{equation}
                \label{Ourloss}
                \mathcal{L}_{R}(\theta) + \alpha {\cal L}_B (\theta),
            \end{equation}
            where $\alpha \in \mathbb{R}$ is a parameter balancing the two losses.
        }

        \modified{It is worth noting} that our boundary loss uses ground-truth boundary information via pre-computed level-set function $\phi_G(q)$, which encodes the distance between each point $q$ and $\partial G$. In Eq. \eqref{boundary-loss}, the softmax for each point $q$ is weighted by the distance function. Such distance-to-boundary information is omitted in widely used regional losses, where all the points within a given region are treated equally, independently of their distances from the boundary.

        \modified{
            Notice that  the global minimum (smallest possible value) of our boundary loss \eqref{boundary-loss} is
            reached when all the negative values in the distance function are included in the sum (i.e., the softmax predictions for the pixels within the ground-truth foreground are equal to 1) and all the positive values are omitted (i.e., the softmax predictions within the background are equal to zero). This means that the global optimum is reached for softmax predictions that correspond exactly to the ground truth, which confirms the meaningfulness of our boundary loss.
            It is also worth noticing that the gradient of our loss is $\phi_G$ multiplied the gradient of the softmax predictions.
            This results in negative factors for the pixels in $G$, which encourages $s_{\theta}$ to increase during SGD, with the magnitude (strength) of the factors depending on the distance between the pixel and the ground-truth boundary (the further the pixel from the boundary, the higher the magnitude of the factor). Positive factors for pixels within the background ($\Omega \setminus G$) encourage the softmax predictions to decrease.

          As we will see in our experiments, it is important to use our boundary loss in conjunction with a regional loss
          for the following technical facts. As discussed earlier, the global optimum of our boundary loss corresponds to a strictly negative value, with the softmax probabilities yielding a non-empty foreground region.
          However, an empty foreground, with approximately null values of the softmax probabilities almost everywhere, corresponds to very low gradients. Therefore, this trivial solution is close \revision{to} a local minimum or a saddle point. This is why we integrate our boundary loss with a regional loss: the regional loss guides training during the first epochs and avoids getting stuck in such  trivial solutions. In the next section, we will discuss various scheduling strategies for updating the weight of the boundary loss during training, with the boundary loss becoming very dominant, almost acting alone, towards the end of the training process. It is also worth noting that this behaviour of boundary terms is conceptually similar to the behaviour of classical and popular contour-based energies for segmentation, e.g., level set Geodesic Active Contours (GAC) \cite{caselles-97} or discrete Markov Random Fields (MRFs) for boundary regularization and edge alignment \cite{Boykov06}, which require additional regional terms to avoid trivial empty-region solutions.
        }

    \section{Experiments}
        \modified{
            In this section, we perform two sets of experiments.
            First, we perform comprehensive evaluations demonstrating to positive effect of integrating our boundary loss with different regional losses $\mathcal{L}_R$.
            Then, we perform a study on the different strategies for selecting and scheduling weight $\alpha$ in \eqref{Ourloss}, showing its impact on performances and good default values for new applications.
        }

        \subsection{Datasets}
            To evaluate the proposed boundary loss, we selected two challenging brain lesion segmentation tasks, each corresponding to highly unbalanced classes.

            \paragraph{ISLES:} The training dataset provided by the ISLES organizers is composed of 94 ischemic stroke lesion multi-modal scans. In our experiments, we split this dataset into training and validation sets containing 74 and 20 examples, respectively. Each scan contains Diffusion maps (DWI) and Perfusion maps (CBF, MTT, CBV, Tmax and CTP source data), as well as the manual ground-truth segmentation. \revision{The spatial resolution goes from $0.8mm\times0.8mm\times4mm$ to $1mm\times1mm\times12mm$.} More details can be found in the ISLES website\footnote{\url{http://www.isles-challenge.org}}.

            \paragraph{WMH:} The public dataset of the White Matter Hyperintensities (WMH)\footnote{\url{ http://wmh.isi.uu.nl}} MICCAI 2017 challenge contains 60 3D T1-weighted scans and 2D multi-slice FLAIR acquired from multiple vendors and scanners in three different hospitals. \revision{The spatial resolution goes from $0.95mm\times0.95mm\times3mm$ to $1.21mm\times1mm\times3mm$ for each volume.} In addition, the ground truth for the 60 scans is provided. From the whole set, 50 scans were used for training, and the remaining 10 for validation.

            % Ratios for the training/validation sets:

            % WMH:
            % Ratio between pos/neg: 0.0022208299628979572 (311351/140195785)
            % Ratio between pos/neg: 0.001577004574422543 (43929/27855975)

            % ISLES:
            % Ratio between pos/neg: 0.012254338267507232 (331632/27062416)
            % Ratio between pos/neg: 0.014654271657429144 (79507/5425517)

        \modified{
            \subsection{Compared losses}
                As stated previously, our proposed boundary loss can be combined with any standard regional loss.
                In the following experiments, we evaluated different popular ones:

                \paragraph{\textbf{GDL}}\cite{sudre2017generalised}
                    We use the binary case of this loss, described in Equation \eqref{eq:gdl}. This is also the baseline loss that we use for the experiments on the selection of $\alpha$. An important advantage of this loss is that it is hyper-parameter free.
                \paragraph{\textbf{Distance weighted cross-entropy}}\cite{ronneberger2015u}
                    UNet original paper proposed this loss as a way to integrate spatial information during the training.
                    It is a modified weighted cross-entropy loss, where the weight for each pixel depends both on the class distribution, and its distance to the two cells closest boundaries.
                    We adapted it for our case, where we take into account only one distance:
                    \[ \mathcal{L}_{\text{UNET}}(\theta) = - \int_C \int_\Omega u_c(p)\log s_\theta^c(p)dpdc , \]
                    where $C$ is the set of classes and $s_\theta^c(p)$ are the network predictions for class $c$. $u_c(p)$ is defined as:
                    \[ u_c(p) = g_c(p) \left[ w_c + w_0e^{\frac{- D_G(p)^2}{2 \sigma^2}} \right], \]
                    where $w_c = \frac{\int_\Omega g_c(p)dp}{|\Omega|}$, and $w_0=10$ and $\sigma=5$ are two hyper-parameters. We kept the paper's default values.

                \paragraph{\textbf{Focal loss}}\cite{lin2018focal}
                    The idea of this loss is to give hard examples a more important weight:
                    \[ \mathcal{L}_{\text{FOCAL}} = - \int_C \int_\Omega (1 - s_\theta^c(p))^\gamma g_c(p) \log{s_\theta^c(p)} dpdc , \]
                    with $\gamma=2$ as default hyper-parameter.
                    Therefore, during training, pixels correctly classified with a high confidence will have little to no influence.
        }

        \revision{
            \paragraph{\textbf{Hausdorff loss}}\cite{karimi2019reducing}
                This closely related loss is also designed to minimize some distance between the two boundaries, but through a different path. We refer to this loss as $\mathcal{L}_{HD}$.

                \[ \mathcal{L}_{HD} = \int_\Omega (g(p) - s_\theta(p))^{2} (D_G(p)^{\beta} + D_S(p)^{\beta})    dp , \]
                where $D_S$ denotes the distance function from predicted boundary $S$, after thresholding $s_\theta$. $\beta$ is a hyper-parameter, which the authors of \cite{karimi2019reducing} set to 2 following a grid search. Unlike our boundary loss, computing $D_S$ cannot be done in a single step before training. The distance needs to be re-computed at each epoch during training, for all the images. It also requires to store the whole volume $\Omega$ in memory, as we cannot compute the distance map for only a subset of $\Omega$. These might be important computational and memory limitations, more so when dealing with large images, as is the case for 3D distance maps.
        }

        \revision{
            \subsection{2D and 3D distance maps}
                While the main experiments resort to a distance map computed from each individual 2D slice, we evaluate the proposed boundary loss with a distance map computed from the whole initial 3D segmentation mask. Equation \eqref{boundary-loss} enables us to have only a subset of $\Omega$ at each update, making it possible to use a 3D distance map with mini-batches of 2D slices.}

        \modified{
            \subsection{Selection of $\alpha$}
                We study several strategies for selecting $\alpha$, and its effect on the performances. On top of a constant pre-selected $\alpha$, we evaluated simple scheduling strategies to update it during the training.

                \paragraph{\textbf{Constant $\alpha$}} The simplest method would be to use a constant during the whole training, but this might require careful tuning of its value.

                \paragraph{\textbf{Increase $\alpha$}} We start with a low value of $\alpha > 0$, and increase it gradually at the end of each epoch. The weight of the regional loss $\mathcal{L}_R$ remains constant over time. At the end of the training, the two losses have the same weight.

                \paragraph{\textbf{Rebalance $\alpha$}} First we rewrite our combined loss as $(1 - \alpha) \mathcal{L}_R + \alpha \mathcal{L}_B$. As for the increase strategy, we start with a low $\alpha >0$, and increase it over time.
                In this way, we give more importance to the regional loss term at the beginning while gradually increasing the impact of the boundary loss term.
                Note that we make sure that the weight for $\mathcal{L}_R$ never reaches 0; the two losses are used at all times during training.
        }

        \subsection{Implementation details}
            \paragraph{\textbf{Data pre-processing}} While the scans are provided as 3D images, we process them as a stack of independent 2D images, which are fed into the network. In fact, the scans in some datasets, such as ISLES, contain between 2 and 16 slices, making them ill-suited for 3D convolutions in those cases.
            The scans were normalized between 0 and 1 before being saved as a set of 2D matrices, and re-scaled to 256$ \times $256 pixels if needed.
            When several modalities were available, all of them were concatenated before being used as input to the network. We did not use any data augmentation in our experiments.

            \paragraph{\textbf{Architecture and training}} We employed UNet \cite{ronneberger2015u} as deep learning architecture in our experiments. To train our model, we employed Adam optimizer, with a learning rate of $0.001$ and a batch size equal to 8. The learning rate is halved if the validation performances do not improve during 20 epochs. We did not use early stopping. %All the experiments were run twice.

            To compute the level set function $\phi_G$ in Eq. (\ref{boundary-loss}), we used standard SciPy functions\footnote{\url{https://docs.scipy.org/doc/scipy-0.14.0/reference/generated/scipy.ndimage.morphology.distance_transform_edt.html}}. Note that, for slices containing only the background region, we used a zero-distance map, assuming that the \modified{regional loss} is sufficient in those cases.
            \modified{For the increase and rebalance $\alpha$ scheduling strategies, we start with $\alpha = 0.01$ and increase it by $0.01$ at the end of each epoch.
                For all the experiments comparing different losses, we use the same rebalance strategy, with the same hyper-parameters.
            }
            In addition, we evaluated the performance when the boundary loss is the only objective, i.e., $\alpha = 0$.

            For our implementation, we used PyTorch \cite{paszke2017automatic}, and ran the experiments on a machine equipped with an NVIDIA GTX 1080 Ti GPU with 11GBs of memory. Our code (data pre-processing, training and testing scripts) is publicly available\footnote{\url{https://github.com/LIVIAETS/surface-loss}}.
            \revision{As \cite{karimi2019reducing} did not release their code, we relied on the re-implementation from \cite{ma2020distance}\footnote{\url{https://github.com/JunMa11/SegWithDistMap}}.}

            \paragraph{\textbf{Evaluation}} For evaluation purposes, we employ the common Dice Similarity Coefficient (DSC) and modified Hausdorff Distance\footnote{\revision{We report the 95th percentile distance value instead of the maximum-distance value.}} (HD95) metrics.

        \subsection{Results}
            \modified{
                \subsubsection{Comparison of regional losses}
                    In this section, we detail the results that we obtained when using different regional losses $\mathcal{L}_R$.
            }

            \modified{
                \paragraph{\textbf{Quantative evaluation}} Table \ref{tab:quantitative_results} reports the DSC and HD performances for our experiments using four different choices of $\mathcal{L}_R$, with each regional term used either alone or in conjunction with our boundary loss in Eq. (\ref{Ourloss}), on the ISLES and WMH datasets.
                In most of the settings, adding the boundary loss during training improves the performances, as reflected in the significantly better DSC and HD values. For instance, on the ISLES segmentation task, adding the boundary loss yielded about $13\%$ improvement in DSC over using Generalized Dice loss alone, and about $3\%$ improvement over using UNet cross-entropy or focal loss alone.}
            \revision{The discrepancy of the improvements the boundary loss brings might be due to the difference in the difficulty of the tackled tasks. The more difficult the tasks (i.e., when regional terms have difficulty achieving good performances), the larger the gain boundary loss brings (as it complements regional information).
                GDL/ISLES is a noticeable case, where
                boundary loss corrected substantially the performance of the GDL regional loss, making it the winning competitor (although, without boundary information, it is the worse-performing regional loss).
            }

            \modified{
                The mixed results with the UNet cross-entropy (improvement on ISLES, but stall on WMH), and the difference on the HD95 metrics can potentially be explained by a toxic interplay between the two losses: both of them are trying to use the distance from the boundary information, potentially counter-acting each others, and introducing instability.
            }

            \revision{
                Computing the distance map from the 3D volume rather than from the 2D slices gives a small boost in performance (about 1\% DSC), and is more noticeable on the training curve for WMH (Figure \ref{fig:learning_curves_both}). This difference could be explained by the spacing between the slices on the $z$ axis: they are quite close (and correlated) in the case of WMH.
                However, in the case of ISLES, the big spacing (around 1cm)  makes slices quite independent. Adding 3D information in this case is less helpful.

                While the Hausdorff loss \cite{karimi2019reducing} also improves the results over the GDL alone (around 7\% on ISLES), its performance is not always at the same level as boundary loss (similar performance on WMH, but lower on ISLES). This is consistent with the findings of \cite{ma2020distance}, which found that the differences in performances are dataset dependent.
            }

            \modified{
                \begin{table}[h]
                    % constrained_cnn-190929-ba64b34-feynman-isles_light.tar.gz
                    \centering

                    \footnotesize
                    \begin{tabular}{l|c|c||c|c}
                        \multirow{2}{*}{Loss} & \multicolumn{2}{c||}{ISLES} & \multicolumn{2}{c}{WMH}\\
                         \cline{2-5}
                         & DSC & HD95 (mm) & DSC & HD95 (mm) \\
                        \hline
                        \hline
                        ${\cal L}_B$ & NA & NA & NA  & NA \\
                        ${\cal L}_{HD}$ & NA & NA & 0.638 (NA)  & 4.578 (NA) \\
                        \hline
                        GDL & 0.511 (0.016) & 5.320 (1.742) & 0.768 (0.051) & 3.634 (2.570) \\
                        ~ w/ ${\cal L}_B$ (2D) & 0.644 (0.026) & 4.795 (3.712) & 0.793 (0.006) & 2.039 (1.834) \\
                        ~ w/ ${\cal L}_B$ (3D) & \textbf{0.659 (0.001)} & 2.725 (2.196) & \textbf{0.818 (0.003)} & \textbf{1.702 (1.982)} \\
                        ~ w/ ${\cal L}_{HD}$ & 0.582 (0.015) & 4.126 (1.634) & 0.805 (0.015) & 2.151 (2.100) \\
                        \hline
                        UNet cross-entropy \cite{ronneberger2015u} & 0.608 (0.025) & 4.572 (0.675) & 0.757 (0.015) & 4.355 (3.388) \\
                        ~ w/ ${\cal L}_B$ (2D) & 0.631 (0.016) & 5.961 (2.291) & 0.756 (0.022) & 2.887 (2.629) \\
                        \hline
                        Focal loss \cite{lin2018focal} & 0.631 (0.046) & 4.989 (2.775) & 0.808 (0.026) & 1.816 (1.370) \\
                        ~ w/ ${\cal L}_B$ (2D) & 0.650 (0.019) & \textbf{1.770 (0.549)} & 0.786 (0.031) & 2.258 (2.513) \\
                        \hline
                    \end{tabular}
                    \caption{Average DSC and HD95 values (and standard deviation over three independent runs) achieved on the validation subset. \revision{Best results highlighted in bold.}}
                    \label{tab:quantitative_results}
                \end{table}

                \begin{figure}
                    \centering
                    \includegraphics[width=0.45\textwidth]{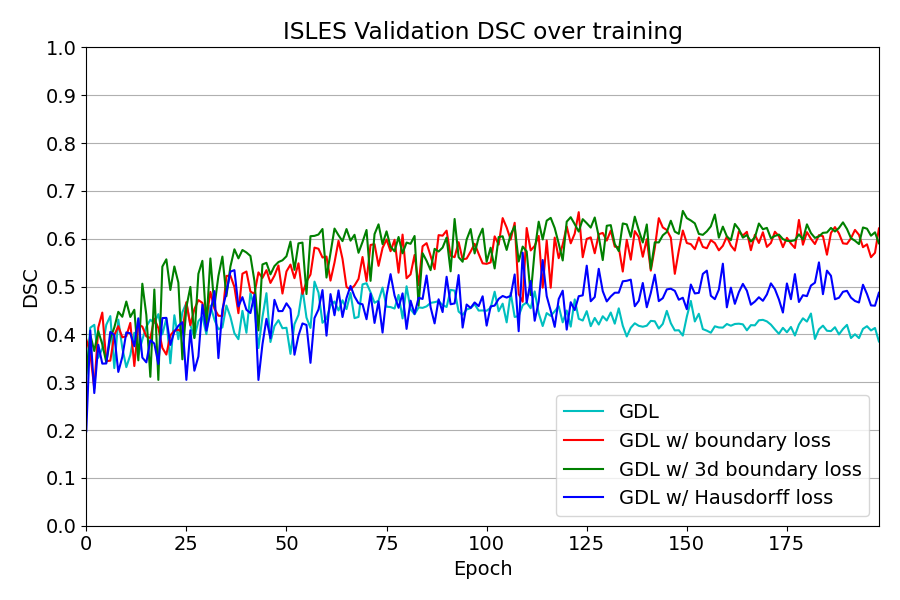}
                    \includegraphics[width=0.45\textwidth]{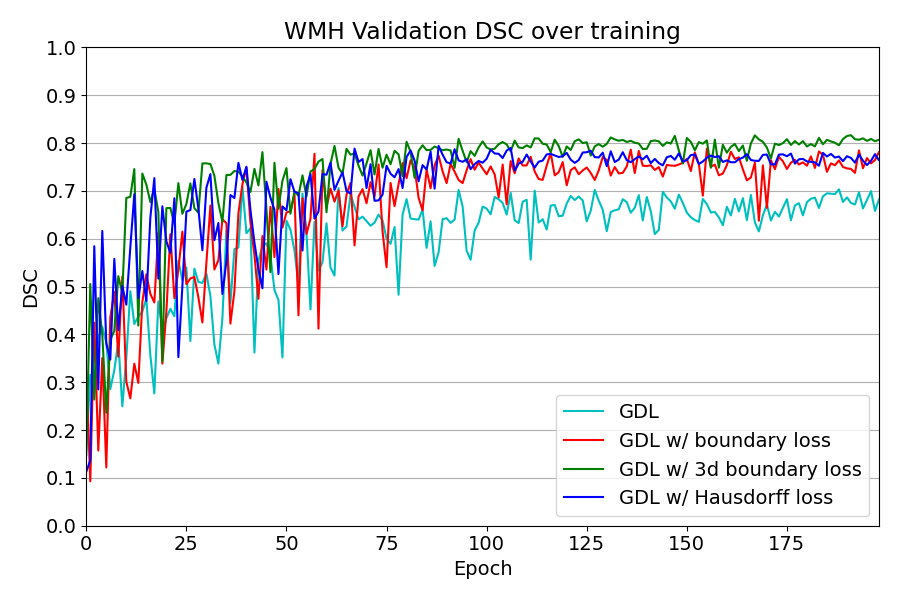}

                    \includegraphics[width=0.45\textwidth]{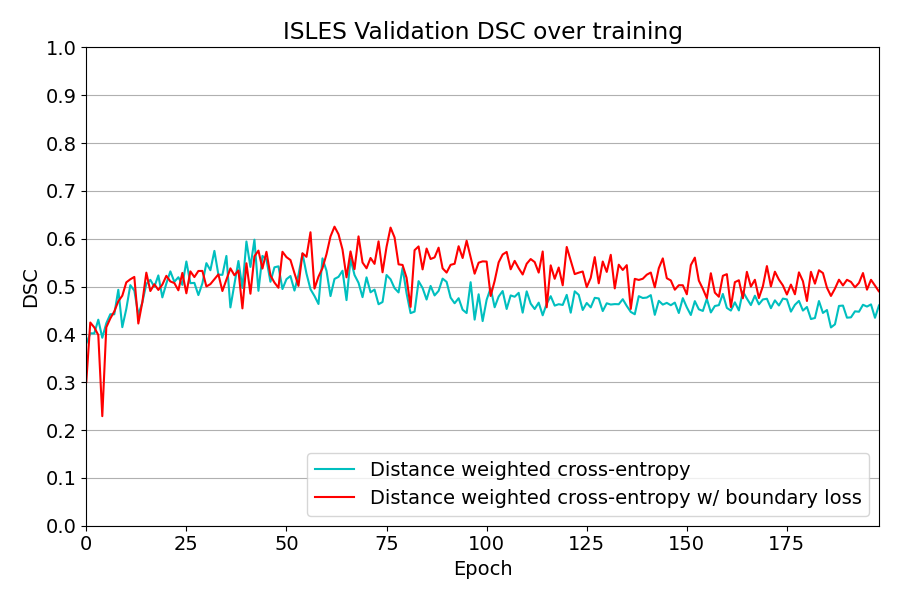}
                    \includegraphics[width=0.45\textwidth]{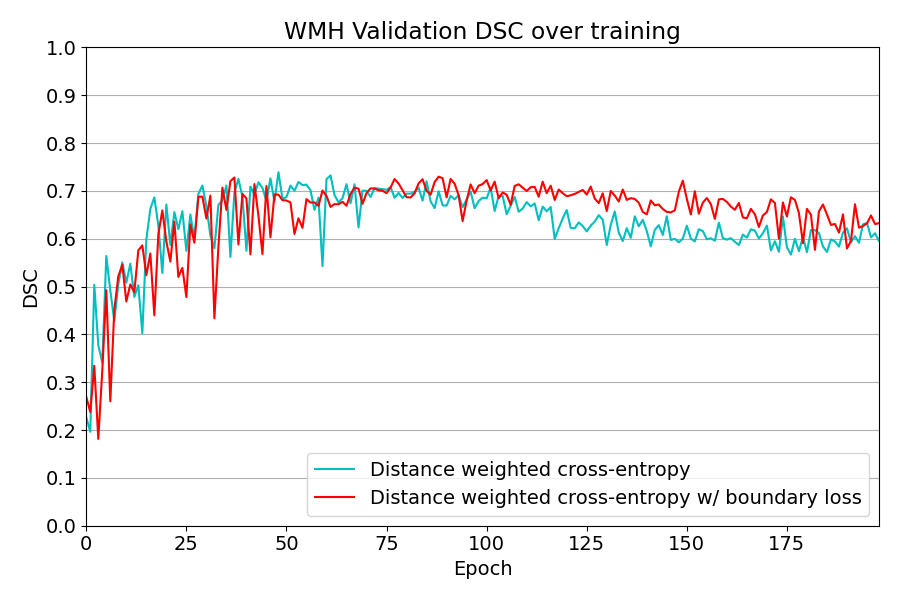}

                    \includegraphics[width=0.45\textwidth]{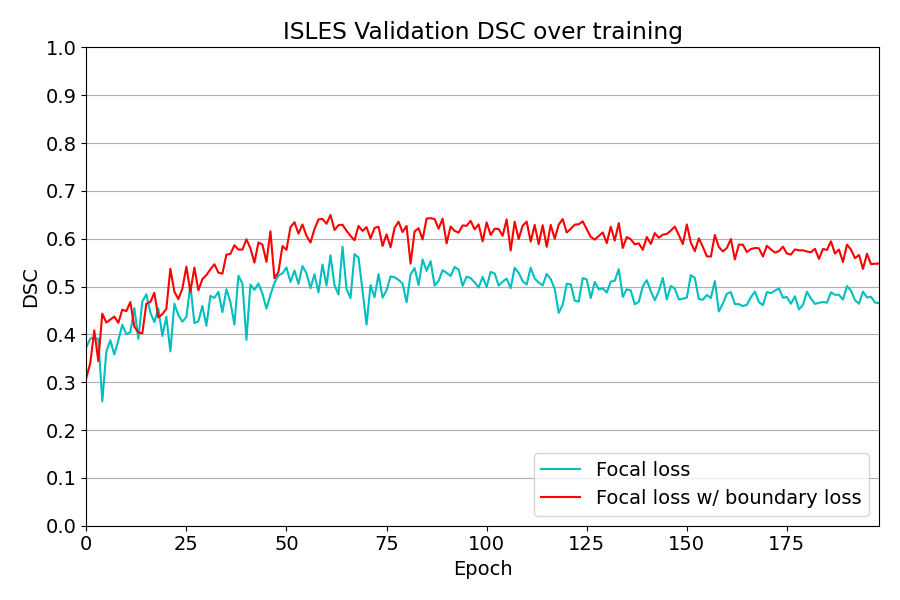}
                    \includegraphics[width=0.45\textwidth]{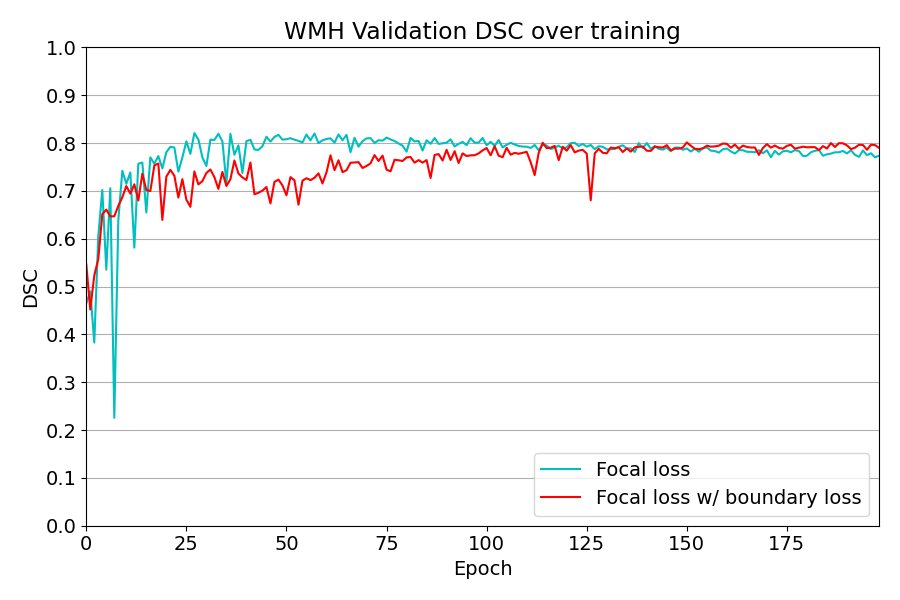}

                    \caption{Evolution of DSC values on validation subsets, for different base losses, on both ISLES and WMH. Best viewed in colors.}
                    \label{fig:learning_curves_both}
                \end{figure}
            }

            Using the boundary loss alone does not yield the same competitive results as a joint loss (i.e., boundary and region), making the network collapse quickly into empty foreground regions, i.e., softmax predictions close to zero\footnote{For this reason, we do not report metrics in this case, as it would be meaningless.}.
            We believe that this is due to the following technical facts. In theory, the global optimum of the boundary loss corresponds to a negative value, as a perfect overlap sums only over the negative values of the distance map.
            In this case, the softmax probabilities correspond to a non-empty foreground. However, an empty foreground (null values of the softmax probabilities almost everywhere) corresponds to low gradients. Therefore, this
            trivial solution is close a local minimum or a saddle point.
            This is not the case when we use our boundary loss in conjunction with a regional loss, which guides the training during the first epochs and avoids getting stuck in such a trivial solution. The scheduling method then increases the weight of the boundary loss, with the latter becoming very dominant towards the end of the training process.
            This behaviour of boundary terms is conceptually similar to the behaviour of classical and popular contour-based energies for level set segmentation, e.g., geodesic active contours \cite{caselles-97}, which also require additional regional terms to avoid trivial solutions (i.e., empty foreground regions).

            \modified{
                The learning curves depicted in
                % Figures \ref{fig:learning_curves_isles} and \ref{fig:learning_curves_wmh}
                Figure \ref{fig:learning_curves_both}
                show the gap in performances between using a regional loss $\mathcal{L}_R$ alone and when augmented with our boundary loss, for different choices of $\mathcal{L}_R$.
                In most of the settings, the difference becomes significant at convergence.
                This behaviour is most visible when $\mathcal{L}_R = \mathcal{L}_{\text{GDL}}$, and is consistent for both metrics and both dataset, which clearly shows the benefits of employing the proposed boundary loss term.
            }

\begin{figure}[ht]
                \centering
                \includegraphics[width=.95\textwidth]{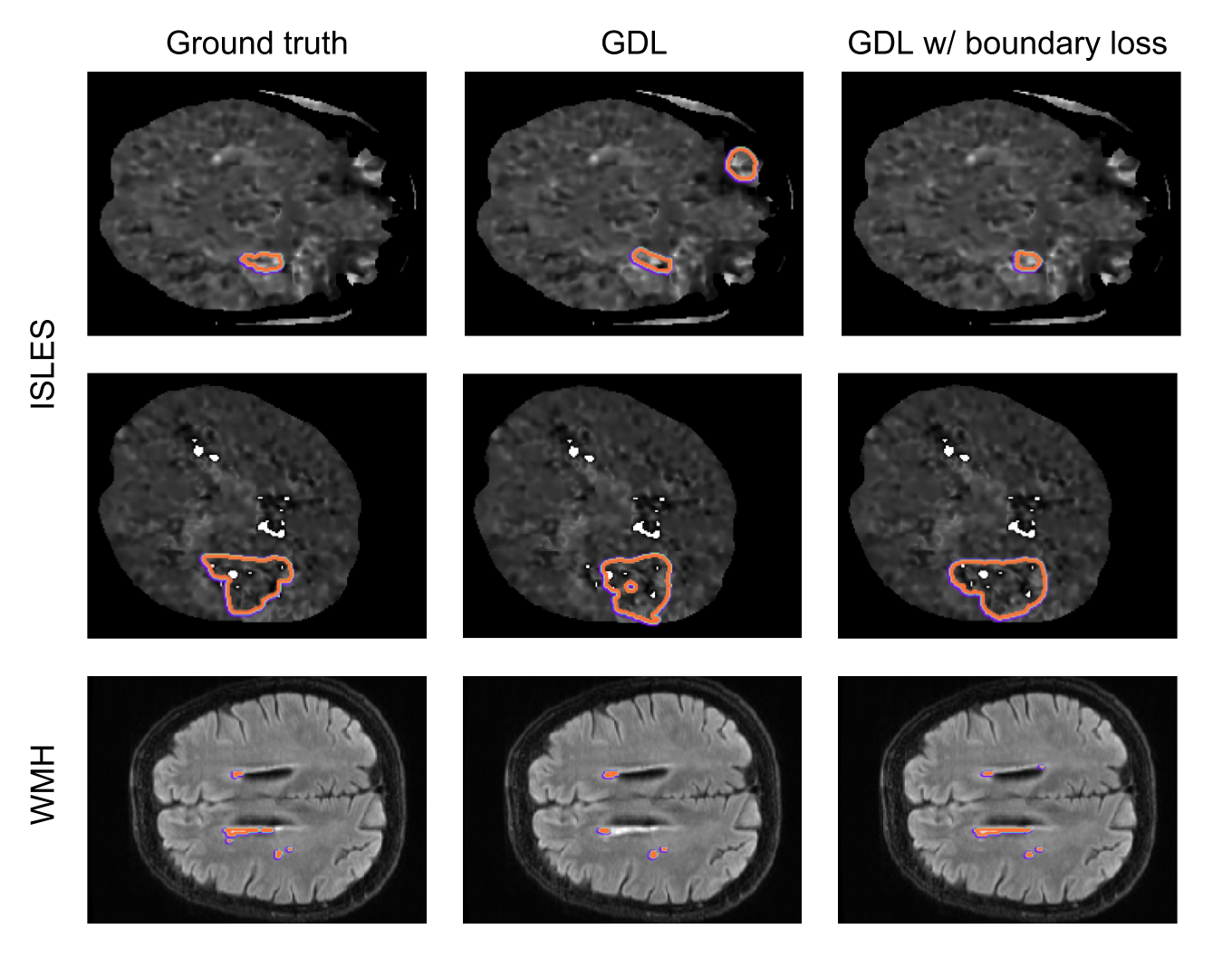}
                \caption{Visual comparison on two different datasets from the validation set.}
                \label{fig:visual_cmp}
            \end{figure}

\begin{figure}
                % constrained_cnn-190929-ba64b34-feynman-isles_light.tar.gz
                \centering
                \includegraphics[angle=90,height=1\textheight]{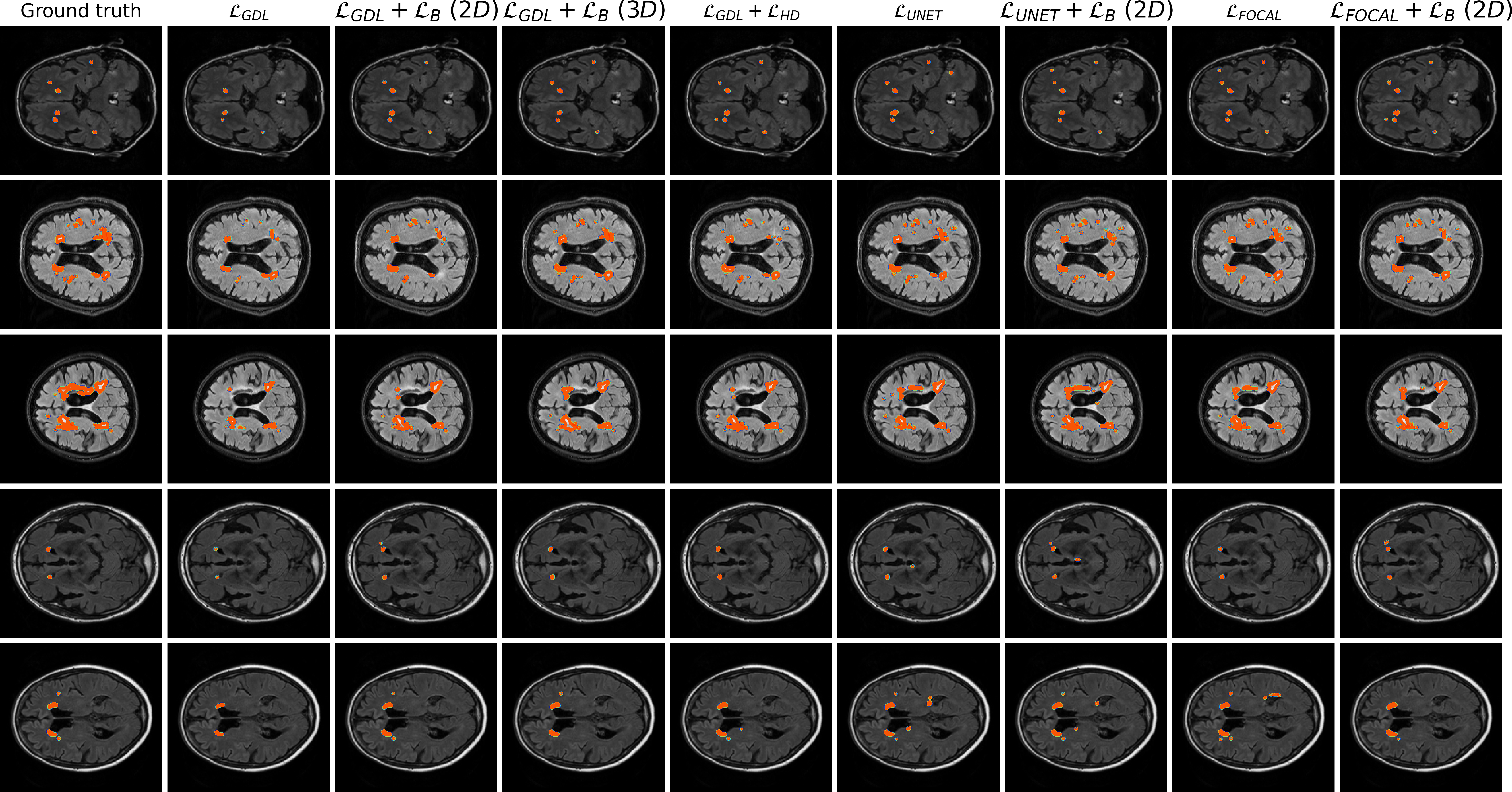}
                \caption{\modified{Visual comparison on the WMH dataset for different training losses. The last column depicts a failure case, where the proposed loss does not enhance the regional loss performance. Best viewed in colors.}}
                \label{fig:wmh_visual}
            \end{figure}

\begin{figure}
                % constrained_cnn-190929-ba64b34-feynman-isles_light.tar.gz
                \centering
                \includegraphics[angle=90,height=1\textheight]{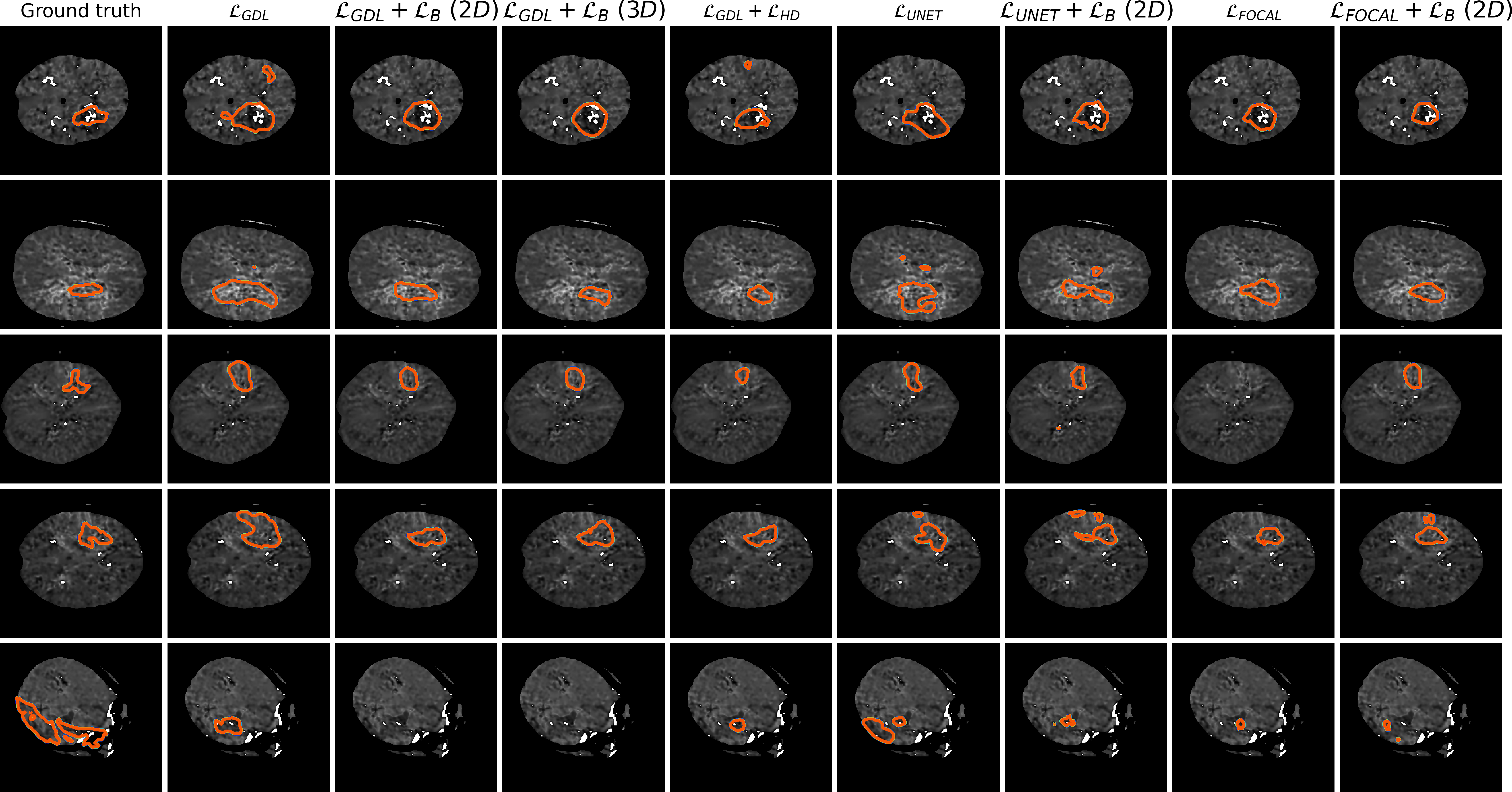}
                \caption{\modified{Visual comparison on the ISLES dataset for different training losses. The last column depicts a failure case, where the proposed loss does not enhance the regional loss performance. Best viewed in colors.}}
                \label{fig:isles_visual}
            \end{figure}

            \paragraph{\textbf{Qualitative evaluation}} Qualitative results are depicted in Fig. \ref{fig:visual_cmp}. Inspecting these results visually, we can observe that there are two major types of improvements when employing the proposed boundary loss. %: detection of small regions and reduction of false positives.
            First, as the methods based on DSC losses, such as GDL, do not use spatial information, prediction errors are treated equally. This means that the errors for pixels/voxels in an already detected object have the same importance as the errors produced in completely missed objects. On the contrary, as our boundary loss is based on the distance map from the ground-truth boundary $\partial G$, it will penalize much more such cases, helping to recover small and far regions. This effect is best illustrated in Fig. \ref{fig:gt_gdl_surface} and Fig. \ref{fig:visual_cmp} (third row). False positives (first row in Fig. \ref{fig:visual_cmp}) will be far away from the closest foreground, getting a much higher penalty than with the GDL alone. This helps in reducing the number of false positives.
            \revision{Additional qualitative results for other base losses, and their combination with the proposed boundary loss, are depicted in Figures \ref{fig:wmh_visual}, \ref{fig:isles_visual}. These figures also show failure cases (\textit{last column}) of the boundary loss.}


            \paragraph{\textbf{Computational complexity}} It is worth mentioning that, as the proposed boundary loss term involves an element-wise product between two matrices -- i.e., the pre-computed level-set function $\phi_G$ and the softmax output $s_{\theta}(p)$ -- the complexity that it adds is negligible \revision{as showed in Table \ref{tab:computation_speed}.}
            \revision{Contrary, the Hausdorff loss \cite{karimi2019reducing} introduces around 10\% of slowdown in the training process. This will be further magnified if we generalize to multi-class problems, where an individual distance map should be computed for each class.}

                \begin{table}
                    \centering

                    \revision{
                    \begin{tabular}{l|c|c}
                        \multirow{2}{*}{Loss} & \multicolumn{2}{c}{Time (s) per batch} \\
                         \cline{2-3}
                         & ISLES  (Batch size = 4) & WMH (Batch size = 8) \\
                        \hline
                        \hline
                        GDL & 0.187 (0.129) & 0.345 (0.132) \\
                        ~ w/ ${\cal L}_B$ & 0.190 (0.128)  & 0.345 (0.129) \\
                        ~ w/ ${\cal L}_{HD}$ & 0.210 (0.108)  & 0.392 (0.092) \\
                    \end{tabular}
                    }
                    \caption{\revision{Training time required by different losses. We report the average and standard deviation batch time in seconds for each method.}}
                    \label{tab:computation_speed}
                \end{table}

            \modified{
                \subsubsection{Selection of $\alpha$}
                    Table \ref{tab:alpha} reports the performances of the proposed approach on the  ISLES  segmentation task for different $\alpha$ values and scheduling techniques.
                    Figure \ref{tab:alpha} shows a subset of the learning curves related to $\alpha$ selection strategies in Table \ref{tab:alpha}.
                    \revision{This is an indication that, while our boundary loss can benefit from a tuned balance between the two losses, even a sub-optimal $\alpha$ can already provide improvement over the regional loss alone.}
                    Observe that increasing the weight of constant $\alpha$ yields better performances, up to a certain value, with the performances decreasing starting from $\alpha=1.5$. With $\alpha=2$,
                    the performance is similar to a network trained with the boundary loss alone. In contrast, using any of the two proposed scheduling strategies (increasing $\alpha$ or re-balancing) yields better results than any constant $\alpha$, without having to explore many configurations.

                    From the learning curves (Figure \ref{fig:alpha}), we can notice that the GDL alone and the GDL with a small constant $\alpha=0.001$ have a similar training DSC over time, but that their validation DSC are significantly different.
                    A similar behaviour can be observed by examining the results with constant $\alpha=1$ and the rebalanced $\alpha$: while the rebalancing training DSC is slightly higher during the whole training, the validation DSC becomes significantly better around half the training time, where the high constant $\alpha$ performances starts decreasing.

                    \begin{table}[h]
                        % constrained_cnn-190929-ba64b34-feynman-isles_light.tar.gz
                        \centering
                        \begin{tabular}{l|l|c|c}
                            \multicolumn{2}{c|}{\multirow{2}{*}{Strategy}} & \multicolumn{2}{c}{ISLES} \\
                            \multicolumn{2}{c|}{} & DSC & HD95 \\
                             \hline
                             \hline
                             \multicolumn{2}{c|}{GDL only} & 0.511 (0.016) & 5.320 (1.742) \\
                             \hline
                             \multirow{7}{*}{Constant $\alpha$} & $\alpha=0.001$ & 0.545 (0.020) & 4.778 (1.546) \\
                              & $\alpha=0.01$ & 0.566 (0.019) & 5.052 (1.395) \\
                              & $\alpha=0.05$ & 0.606 (0.015) & 5.326 (1.712) \\
                              & $\alpha=0.1$ & 0.605 (0.010) & 5.762 (1.782) \\
                              & $\alpha=0.5$ & 0.604 (0.006) & 9.234 (10.463) \\
                              & $\alpha=1$ & \textit{0.628} (0.023) & \textit{2.462} (0.706) \\
                              & $\alpha=1.5$ & 0.565 (0.074) & 3.335 (1.164) \\
                              & $\alpha=2$ & 0.549 (0.084) & 20.275	(16.603) \\
                            \hline
                            \multicolumn{2}{c|}{Increase $\alpha$} & \textit{0.622} (0.004) & \textit{4.952} (1.773) \\
                            \hline
                            \multicolumn{2}{c|}{Rebalance $\alpha$} & \textbf{\textit{0.644}} (0.026) & \textbf{\textit{4.795} (3.712)} \\
                        \end{tabular}
                        \caption{Results on ISLES validation set for different $\alpha$.}
                        \label{tab:alpha}
                    \end{table}

                    \begin{figure}[h]
                        \centering
                        \includegraphics[width=0.49\textwidth]{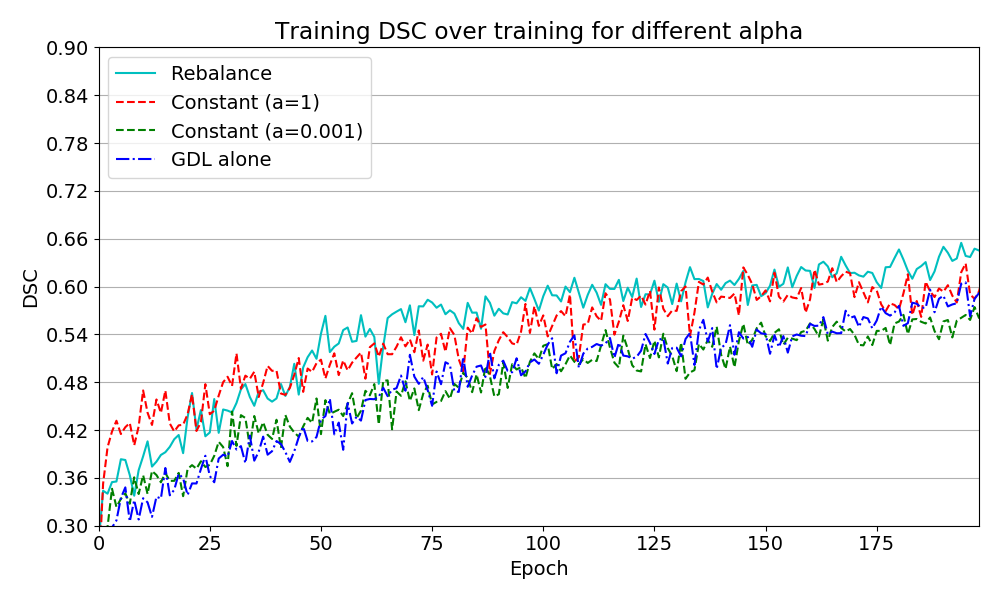}
                        \includegraphics[width=0.49\textwidth]{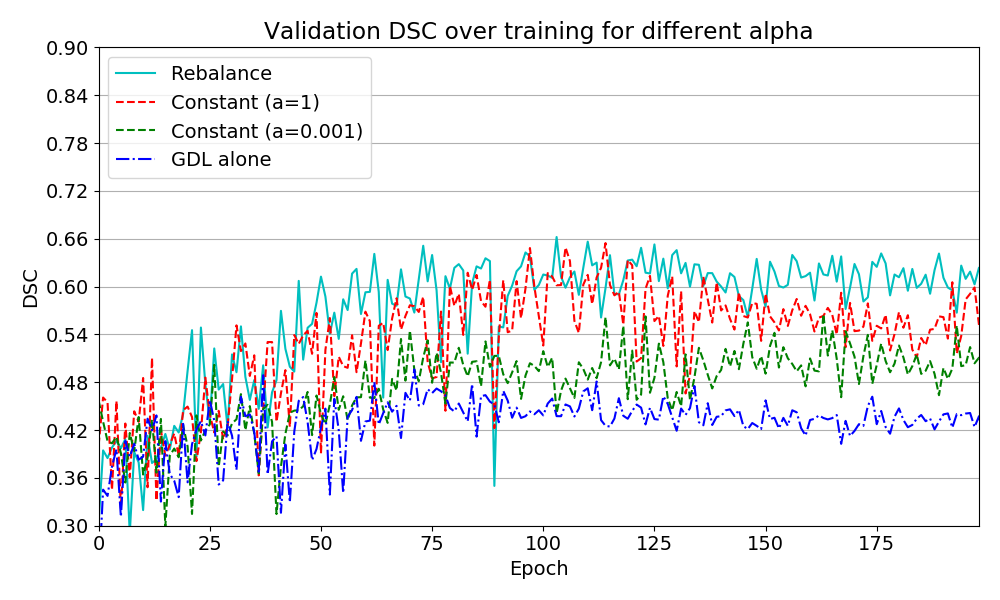}
                        \caption{Comparison of the training and validation DSC curves for different $\alpha$ selection strategies. For readability, not all settings from Table \ref{tab:alpha} have been included. Best viewed in colors.}
                        \label{fig:alpha}
                    \end{figure}

                    \revision{The rebalancing strategy was used in all other experiments, and as showed in Table \ref{tab:quantitative_results}, proved to be a good default strategy to integrate the boundary loss with another regional loss.}
            }

    \section{\revision{Conclusion and future works}}

        We proposed a boundary loss term that can be easily combined with any standard regional loss, to tackle segmentation tasks in highly unbalanced scenarios. Furthermore, the proposed term can be implemented with any existing deep network architecture and for any N-D segmentation problem. Our experiments on two challenging and highly unbalanced datasets demonstrated the benefits of including our boundary loss during training. It consistently improved the performances, and by a large margin on one data set, with enhanced training stability.

        \revision{In this work, we evaluated the proposed boundary loss in the context of class imbalance. However, there are other interesting avenues for extending and evaluating our approach. For instance, our boundary loss has a spatial regularization effect because it is based on distance-to-boundary information. In particular, we observed experimentally that it yield contours, which are, typically, smoother than those obtained with regional losses. Focused on the important problem of unbalanced segmentation, our experiments did not fully investigate the benefits of such a spatial regularization. An interesting future research avenue will be to explore such a regularization effect in applications with challenging imaging noise, which may prevent regional losses from generating smooth contours, e.g., ultrasound imaging. Another limitation of our formulation and experiments is that they were limited to binary (two-region) segmentation problems. It will be interesting to investigate extensions of boundary loss to the multi-region scenario, with competing distance maps from multiple structures and various/complex topological constraints (e.g., one structure fully included within another).}

    \section*{Acknowledgments}
        This work is supported by the National Science and Engineering Research Council of Canada (NSERC), Discovery Grants (DG) program, and by the ETS Research Chair on Artificial Intelligence in Medical Imaging.

    \bibliography{kervadec19}
\end{document}